\documentclass[conference,10pt]{IEEEtran}

\usepackage{amssymb,amsmath,amsfonts,bm,epsfig,graphicx,theorem,latexsym}
\usepackage{rotating,setspace,latexsym,epsf,color,subfigure}
\usepackage{cite,authblk,bbm}
\usepackage[letterpaper, left=1in, right=1in, bottom=1in, top=0.75in]{geometry}

\def \n2{{N_0 \over 2}}

\def \h5{\hspace{0.5in}}

\begin{document}
\IEEEoverridecommandlockouts
\pagestyle{empty}

\title{Trading Off Computation with Transmission in Status Update Systems}

\author{Peng Zou \qquad Omur Ozel \qquad Suresh Subramaniam \\
\normalsize Department of Electrical and Computer Engineering \\
\normalsize George Washington University, Washington, DC 20052 USA \\
\normalsize {\it pzou94, ozel, suresh@gwu.edu} }

\maketitle 

\begin{abstract}
This paper is motivated by emerging edge computing applications in which generated data are pre-processed at the source and then transmitted to an edge server. In such a scenario, there is typically a tradeoff between the amount of pre-processing and the amount of data to be transmitted. We model such a system by considering two non-preemptive queues in tandem whose service times are independent over time but the transmission service time is dependent on the computation service time in mean value. The first queue is in M/GI/1/1 form with a single server, memoryless exponential arrivals, general independent service and no extra buffer to save incoming status update packets. The second queue is in GI/M/1/$2^*$ form with a single server receiving packets from the first queue, memoryless service and a single data buffer to save incoming packets. Additionally, mean service times of the first and second queues are dependent through a deterministic monotonic function. We perform stationary distribution analysis in this system and obtain closed form expressions for average age of information (AoI) and average peak AoI. Our numerical results illustrate the analytical findings and highlight the tradeoff between average AoI and average peak AoI generated by the tandem nature of the queueing system with dependent service times. 
\end{abstract}

\pagestyle{plain}
\setcounter{page}{1}
\pagenumbering{arabic}

\section{Introduction}  

The freshness of the available information coming from continuous data streams is critical in the operation of various Internet-of-Things (IoT) and edge computing applications with examples spanning sensor networking, cognitive radio and vehicular communication networks. A new metric termed age of information (AoI) has found considerable attention in the recent literature as a measure of freshness of available information. After the pioneering works \cite{kaul2012status, kaul2012real} that analyze queuing models motivated from vehicular status update systems, the AoI metric has been found useful in various scenarios. \cite{costa2016age, kam2018age} investigate the role of packet management to decrease the AoI while \cite{inoue2018general} provides a general AoI analysis in various preemptive and non-preemptive queuing disciplines coming after earlier works such as \cite{najm2016age, najm2017status}. References \cite{2018information, bacinoglu2015age, yates2015lazy, wu2017optimal_ieee, arafa2017age, bacinoglu2017scheduling, farazi2018average, bacinoglu2018achieving, feng2018age, arafa2019using} consider AoI in energy harvesting communication systems. Evolution of AoI through multiple hops in networks have been characterized in \cite{bedewy2017age, talak2017minimizing, yates2018age, yates2018status, maatouk2018age} and \cite{kosta2018age, kadota2018scheduling, feng2019adaptive} consider AoI minimization over multiple access and broadcast channels. \cite{Alabbasi2018JointIF} considers scheduling data flows in vehicular communication networks. More recently, \cite{Gong2019ReducingAF,xu2019peak} consider AoI analysis with tandem computing and communication queues. 

\begin{figure}[!t]
\centering{
\hspace{-0.4cm} 
\includegraphics[totalheight=0.095\textheight]{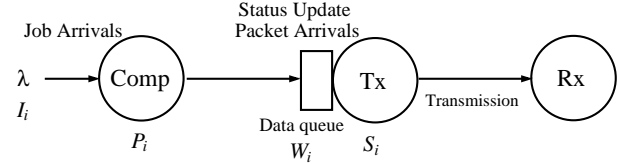}}
\caption{\sl System model with status update packets arriving to a single server transmission queue from the output of a computation server queue in tandem.}\vspace{-0.1in}
\label{fig:1} 
\end{figure}

In this paper, we consider a status update system composed of tandem queues where a computation-type first queue determines status update packets to be sent to a remote monitoring receiver, as shown in Fig. \ref{fig:1}. We are motivated by edge computing applications where packets go through computation and transmission queues in tandem and higher amount of computation before transmission enables a shorter amount of work to be done elsewhere and hence a quicker data transmission. We consider jobs arriving at the computation queue one after the other and enter it only if the server is idle. Any job arriving during a busy cycle of the first queue is discarded and possibly routed to another one. In our work the processing time in the first queue determines the transmission service time necessary for the resulting status update packet. Compared to the existing works \cite{Gong2019ReducingAF,xu2019peak}, our work provides a general analysis with packet management (as opposed to first come first serve) for average AoI and average peak AoI under dependent mean service times and highlights possible tradeoffs between them.  

In this paper, we investigate average AoI and average peak AoI for a tandem non-preemptive queue where jobs arrive at the computation server according to a Poisson process with rate $\lambda$ and processing time $P_i$ for a job has a general distribution as in Fig. \ref{fig:1}. At the end of the job processing, a status update packet to be transmitted is sent to the transmission queue where it waits $W_i$ time units before entering service. This first queue is in the form of M/GI/1/1 as any job arriving to the first queue is discarded right away from the system if the server is busy. The second queue is in the form of GI/M/1/$2^*$ as the arrivals to the second queue come from the first queue with general independent inter-departure times and the service times $S_i$ are exponentially distributed. We use the specific tandem structure of this system to obtain closed form expressions for average peak AoI and average AoI in this system. A crucial aspect of the problem is that the aging starts immediately after job enters the computation server as opposed to the time it enters the transmission queue. This causes the service time in the computation queue to couple the system time and the inter-arrival time in a unique way and our expressions provide explicit dependencies among the parameters in the system. Moreover, in our system model, a functional dependence is assumed between the mean service times in the first queue and the second queue which is typical of edge computing applications. Our numerical results show the benefits obtained by judiciously determining the operating point in AoI and hence understanding the tradeoff between average peak AoI and average AoI generated by the tandem nature of the queueing system. 

\section{The Tandem Queue Model}
\label{sec:Model}

We consider a system with a computation queue followed by a transmission queue as shown in Fig. \ref{fig:1}. In the sequel, we interchangeably refer to the computation and transmission queues as first and second queues. The computation jobs arrive according to a Poisson process with arrival rate $\lambda$. The jobs enter the server only if it is idle and the aging process starts as soon as the computation starts. As soon as the computation is completed, a status update packet, whose length is determined by the duration of computation, is generated and sent to the transmission queue. The transmission system is composed of a transmitter (Tx) and a receiver (Rx) where monitor resides. There is a single data buffer to save the latest arriving packet when the transmission server is busy. The transmitter transmits the status update packets one at a time. The service time is distributed as a memoryless exponential random variable with mean $\mu$, which is dependent on the computation service time. 

This tandem queue model is inspired by packet management schemes in \cite{costa2016age,inoue2018general}. The first queue is in M/GI/1/1 form while the second queue is in GI/M/1/2* form or equivalently non-preemptive last come first serve with discarding. The reason for the second queue to have independent arrivals is that the inter-departure times of the first queue are independent over time and has a general distribution of sum of an exponentially distributed random variable and a general distributed random variable. The tandem nature of these queues deems the resulting problem new and the dependence in service times between them leads to a novel problem that has not been analyzed before to the best of our knowledge. We will perform stationary distribution analysis for average and peak age. The time for a job to be served in the first queue has a general distribution $f_P(p)$, $p \geq 0$, independent of other system variables and independent over time. Corresponding to the general distribution, we have $MGF^{(P)}_{\gamma}$, the moment generating function of the computation time distribution at $-\gamma$ for $\gamma \geq 0$:
\begin{align}
MGF^{(P)}_{\gamma} \triangleq \mathbb{E}[e^{-\gamma P}]
\end{align}
We also use $MGF^{(P,1)}_{\gamma}$ to denote the first derivative of the moment generating function at $- \gamma$.

\subsection{Equivalent Tandem Queue}
\label{sec:eqmodel}

We use an equivalent queue model that yields an identical AoI pattern to our system's. This approach has first appeared in our earlier work \cite{infocom_w, infocom_arxiv} for a single server queue. We adapt this approach to the tandem queue in the current paper by using it in the second queue. In this equivalent model, each arriving packet to the second queue is stored in the queue, the data buffer capacity is unlimited and no packet is discarded. We allow multiple packets to be served at the same time in the second queue. An arriving packet may find the second queue in Idle (Id) or Busy (B) states. If a packet enters the second queue in state (Id), then that packet's service starts right away; otherwise, its service starts after the end of the current service period. The packets arriving to the second queue in state (B) are served together with all other packets that arrive during the same busy period. Note that the first queue remains unchanged in the equivalent queue.

We let $t_i$ denote the time stamp of the event that job $i$ enters the computation queue (we index only those that enter the queue and assume no packet is generated while the computation server is busy), and $t_i'$ the time stamp of the event that the resulting packet $i$ (if selected for service) is delivered to the receiver. Since there is a job and a packet for each index $i$, we use them interchangeably. It is, however, remarkable that the age of the packet is determined with respect to the time it enters the computation server\footnote{This relativity is inherent in multi-hop systems and has been the topic of another work of ours in \cite{isit_arxiv}.}. 

The inter-arrival time between two successive jobs entering the computation queue (i.e., jobs $i-1$ and $i$) is $X_i$. Note that $X_i$ is independent of $X_j$ for $i \neq j$. Each $X_i$ includes a process period of job $i-1$ in the computation queue $P_{i-1}$ and a period of idle time waiting for the next arrival $I_i$; that is, $X_i = I_i + P_{i-1}$ where $I_i$ is independent memoryless exponentially distributed with mean $\frac{1}{\lambda}$. Therefore, we have the moment generating function of inter-arrivals as $MGF^{(X)}_{\gamma}= \frac{\lambda}{\gamma + \lambda}MGF^{(P)}_{\gamma}$.

We also define $T_i$ as the system time for packet $i$ starting from its arrival to the computation queue up until it is delivered to the receiver. We have $T_i = P_i + W_i + S_i$ where $W_i \geq 0$ denotes the length of time packet $i$ spends in the second queue before entering service and $S_i$ is the service time for packet $i$ in the second queue. The instantaneous AoI is the difference of current time and the time stamp of the packet at the receiver: \begin{align} \Delta(t)=t-u(t) \end{align} where $u(t)$ is the time stamp of the latest packet at the receiver at time $t$. We express $u(t)=t_{i^*}$ where $i^*=\max\{i: \ t_i' \leq t\}$.

\begin{figure}[!t]
\centering{
\hspace{-0.5cm} 
\includegraphics[totalheight=0.23\textheight]{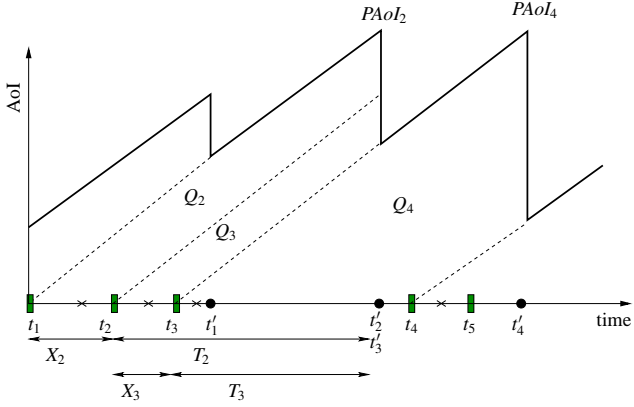}}\vspace{-0.0in}
\caption{\sl AoI evolution for the equivalent tandem queue model with M/GI/1/1 following GI/M/1/2*.}
\label{fig:2} 
\vspace{-0.0in}
\end{figure}

We provide a sample path of the AoI evolution under the equivalent tandem queue model in Fig. \ref{fig:2}. Note that every packet entering the first queue is served in this model. At time $0$, job 1 enters the computation queue while both servers are idle and its computation time is marked as a cross in between $t_1$ and $t_2$. During the system time for packet $1$ in between $t_1$ and $t_1'$, jobs 2 and 3 arrive and service of both jobs in the computation queue end before the service of packet 1 in the transmission queue. Therefore, both packets are kept in the buffer of the second queue to be served together next right after $t_1'$ and the end of service times $t_2'$, $t_3'$ coincide as in Fig. \ref{fig:2}. Note that in the actual system, only packet 3 is served and packet 2 is discarded. At time $t_2'$, the system enters idle state, and packet 4 finds both servers idle. At $t_5$, packet 5 arrives while service of packet 4 continues in the second queue.

We define the areas $Q_i$ under the triangular regions of the AoI curve in the same order as the indices of packets entering the computation queue as shown in Fig. \ref{fig:2}. These definitions are identical to those in \cite{kaul2012real} for first come first serve queuing. Recalling the definition of $X_i$ as the length of time interval between the arrivals of packets $i-1$ and $i$ into the computation server and $T_i$ as the system time for packet $i$ in the equivalent queuing model, we have the average AoI:
\begin{equation}\label{aaoi}
\mathbb{E}[\Delta]=\widetilde{\lambda}\left(\mathbb{E}[XT]+\frac{\mathbb{E}[X^{2}]}{2}\right)
\end{equation}
where $\widetilde{\lambda}$ is the effective arrival rate for the system. In particular, we have
\begin{align}
\widetilde{\lambda} = \frac{\lambda}{ \lambda \mathbb{E}[P] + 1}
\end{align}
and we have
\begin{align}\label{ap1}
\mathbb{E}[X^2]=\mathbb{E}[P^2] + \frac{2 \mathbb{E}[P]}{\lambda} + \frac{2}{\lambda^2}
\end{align}
We also calculate average peak AoI. The peak AoI occurrences are shown in Fig. \ref{fig:2}. In particular, $PAoI_{i^*}$ is the maximum $X_j+T_j$ among all packets $j$ served during a service period and $i^*$ is the smallest index among all of them. In the figure, packets $2$ and $3$ are served together and the peak AoI is $X_2 + T_2$. We assume the system is ergodic and we work with generic variables for inter-arrival time $X$, system time $T$ and $PAoI$ for the maximum $X_j + T_j$ among those that are served together. 

\subsection{Functional Dependence of Mean Service Times}

In our model, we assume that the mean service time of computation queue $\mathbb{E}[P]$ and mean service time of transmission queue $\mathbb{E}[S]=\frac{1}{\mu}$ are dependent through a monotone decreasing function $g$ as:
\begin{align}
\frac{1}{\mu} = g(\mathbb{E}[P])
\end{align}
This dependence reflects the characteristic of computation server and the transmission server in terms of the time it takes to process jobs and packets, respectively. We are motivated by edge computing applications where some computation is performed at the transmitting device to reduce the amount of data to be transferred to a remote computer. Under a fixed transmission rate, expected transmission time is proportional to the length of status update packet which is inversely proportional to the computation time. This operation could also be viewed as compression where the content of data transmission is reduced by removing the noise in the measurements partially or fully before transmission. Since a longer computation time leads to smaller packets, the function $g(.)$ is chosen to be monotone decreasing.

\section{Evaluating Average AoI and Average Peak AoI}
\label{sec:eval}

In this section, we evaluate the AoI in the equivalent tandem queue model. Let us define the state of the second queue packet $i$ finds when it enters the queue as $K_i$, which can take (Id) and (B) states. We note that $K_i$ is a two-state Markov chain. Conditioned on $K_{i-1}=(Id)$, $K_{i}=(Id)$ only if $I_i + P_i > S_{i-1}$. Similarly, conditioned on $K_{i-1}=(B)$, $K_{i}=(Id)$ only if $I_i + P_i > W_{i-1} + S_{i-1}$ where $W_i$ denotes residual service time, which is also the waiting time for packet $i$ in the second queue conditioned on $K_i=(B)$. Note that both $W_i$ and $S_i$ are exponentially distributed with mean $\frac{1}{\mu}$ and they are independent variables. This generates a two-state Markov chain with transition probabilities:
\begin{align}
\mbox{Pr}[K_{i}=(B)|K_{i-1}=(Id)]&=\mbox{Pr}[I_i + P_i < S_{i-1}]\\
\mbox{Pr}[K_{i}=(Id)|K_{i-1}=(B)]&=\mbox{Pr}[I_i + P_i > W_{i-1} + S_{i-1}]
\end{align} 
We calculate these probabilities as
\begin{align}
\mbox{Pr}[I_i + P_i < S_{i-1}] &= \mathbb{E}[e^{-\mu (I_i + P_i)}] =\frac{\lambda}{\lambda + \mu} MGF_{\mu}^{(P)} 
\end{align}
\begin{align}\nonumber
\mbox{Pr}&[I_i + P_i < W_{i-1} + S_{i-1}] \\ &= \mathbb{E}[e^{-\mu (I_i + P_i)} + \mu (I_i + P_i)e^{-\mu (I_i + P_i)}] \\ &=  \frac{\lambda(\lambda + 2\mu)}{(\lambda + \mu)^2}MGF_{\mu}^{(P)} + \frac{\lambda \mu}{\lambda + \mu} MGF_{\mu}^{(P,1)} 
\end{align}
and $\mbox{Pr}[I_i + P_i > W_{i-1} + S_{i-1}] = 1 - \mbox{Pr}[I_i + P_i < W_{i-1} + S_{i-1}]$. Then, the stationary probabilities are
\begin{align}
p_B=\frac{\lambda (\lambda + \mu) MGF_{\mu}^{(P)}}{(\lambda+\mu)^2-\lambda \mu MGF_{\mu}^{(P)} - \lambda \mu (\lambda + \mu) MGF_{\mu}^{(P,1)}} \label{pb}
\end{align}
and $p_I = 1- p_B$ where we define $p_B=\mbox{Pr}[K_i = (B)]$.  

\subsection{Average AoI}

In this subsection, we evaluate $\mathbb{E}[XT]$ and put it in (\ref{aaoi}) along with (\ref{ap1}) to get a closed form expression for average AoI. We next treat the two conditions $K_{i-1}=(Id)$ and $K_{i-1}=(B)$.

\subsubsection{$\mathbb{E}[X_iT_i \ | \ K_{i-1}=(Id)]$}

In this case, packet $i-1$ finds the second queue in (Id) state. $X_i = P_{i-1} + I_{i}$ and if $I_i + P_i > S_{i-1}$, then $T_i = P_i + S_i$. If $I_i + P_i < S_{i-1}$, then $T_i = P_i + W_i + S_i$ where $W_i$ is the residual service time observed by packet $i$ before entering the transmission server. We evaluate the conditional expectation as:
\begin{align*}
\mathbb{E}[X_iT_i|K_{i-1}=(Id)] &= \mathbb{E}[(P_{i-1} + I_i)(P_i + S_i)] \\ &\  +\mathbb{E}[(P_{i-1} + I_i)W_i \mathbbm{1}_{I_i + P_i < S_i}] \\ &= \mathbb{E}[(P_{i-1} + I_i)(P_i + S_i)] \\ &\  +\mathbb{E}[(P_{i-1} + I_i)W_ie^{-\mu (I_i + P_i)}] \\ &= \mathbb{E}^2[P] + \frac{\mathbb{E}[P]}{\lambda} + \frac{1}{\lambda \mu} + \frac{\mathbb{E}[P]}{\mu}  \\ &\ + \frac{\lambda (\lambda + \mu) \mathbb{E}[P] + \lambda}{\mu (\lambda + \mu)^2} MGF_{\mu}^{(P)}
\end{align*}

\subsubsection{$\mathbb{E}[X_iT_i \ | \ K_{i-1}=(B)]$}

In this case, packet $i-1$ finds the second queue in (B) state. $X_i = P_{i-1} + I_{i}$ and if $I_i + P_i > W_{i-1} + S_{i-1}$, then $T_i = P_i + S_i$. If $I_i + P_i < W_{i-1} + S_{i-1}$, then $T_i = P_i + W_i + S_i$. We evaluate the conditional expectation:
\begin{align*}\nonumber
\hspace{-0.0in}\mathbb{E}[X_iT_i|K_{i-1}=(B)] & = \mathbb{E}[(P_{i-1} + I_i)(P_i + S_i)] \\ \nonumber & \hspace{-0.5in} +\mathbb{E}[(P_{i-1} + I_i)W_ie^{-\mu (I_i + P_i)}] \\  &\hspace{-0.5in} + \mathbb{E}[(P_{i-1} + I_i)W_i\mu (I_i + P_i) e^{-\mu (I_i + P_i)}] \\ \nonumber &= \mathbb{E}^2[P] + \frac{\mathbb{E}[P]}{\lambda} + \frac{1}{\lambda \mu} + \frac{\mathbb{E}[P]}{\mu}  \\ \nonumber &\hspace{-0.5in} + \frac{\lambda (\lambda + \mu) \mathbb{E}[P] + \lambda}{\mu (\lambda + \mu)^2} MGF_{\mu}^{(P)} \\ &\hspace{-0.5in} \nonumber +\frac{\lambda \mathbb{E}[P](\lambda + \mu + 1)}{(\lambda + \mu)^2}MGF_{\mu}^{(P)}  \\ &\hspace{-0.5in} +\frac{2 \lambda MGF_{\mu}^{(P)} + \lambda(\lambda + \mu)MGF_{\mu}^{(P,1)}}{(\lambda + \mu)^3}
\end{align*}
We finally use ergodicity of the system and determine $\mathbb{E}[X_iT_i]$:
\begin{align*}\nonumber
\mathbb{E}[X_iT_i]&=\mathbb{E}[X_iT_i|K_{i-1}=(B)]p_{B} \\ \nonumber &\qquad \qquad + \mathbb{E}[X_iT_i|K_{i-1}=(Id)]p_{I} \end{align*} \begin{align*} \nonumber &= \mathbb{E}^2[P] + \frac{\mathbb{E}[P]}{\lambda} + \frac{1}{\lambda \mu} + \frac{\mathbb{E}[P]}{\mu}  \\ \nonumber &\quad + \frac{\lambda (\lambda + \mu) \mathbb{E}[P] + \lambda}{\mu (\lambda + \mu)^2} MGF_{\mu}^{(P)} \\ \nonumber &\quad +p_B \frac{\lambda \mathbb{E}[P](\lambda + \mu) + 2\lambda}{(\lambda + \mu)^3}MGF_{\mu}^{(P)}  \\ &\quad +p_B\frac{ \lambda \mathbb{E}[P] + 1}{\lambda + \mu}MGF_{\mu}^{(P,1)}
\end{align*} 
where $p_B$ is as in (\ref{pb}).

\subsection{Average Peak AoI}

In this subsection, we evaluate $\mathbb{E}[X_{i^*} + T_{i^*}]$ where $i^*$ is the packet index corresponding to the minimum index in a given service period. We have $\mathbb{E}[X_{i^*}+T_{i^*}]=\frac{\mathbb{E}[(X_{i}+T_{i})\mathbbm{1}_{i=i^*}]}{\mbox{Pr}(i=i^*)}$ where $\mathbbm{1}_{i=i^*}$ is the indicator function of whether a given packet is the minimum index in a given service period and $\mbox{Pr}(i=i^*)$ refers to its probability. As before, we will treat two conditions $K_{i-1}=(Id)$ and $K_{i-1}=(B)$ separately for both terms.

\subsubsection{$\mathbb{E}[(X_{i}+T_{i})\mathbbm{1}_{i=i^*} \ | \ K_{i-1}=(Id)]$}

In this case, if $I_i + P_i > S_{i-1}$, then $T_i = P_i + S_i$. If $I_i + P_i < S_{i-1}$, then $T_i = P_i + W_i + S_i$. Conditioned on (Id) state for packet $i-1$, the next packet index $i$ will be the minimum index among all those being served together with certainty. Therefore, we have $\mbox{Pr}(i=i^* \ | \ K_{i-1}=(Id)) = 1$. Additionally, we have
\begin{align}\nonumber
\mathbb{E}[(X_{i}+&T_{i})\mathbbm{1}_{i=i^*} \ | \ K_{i-1}=(Id)]  \\ &\hspace{-0.5in}=  \mathbb{E}[P_{i-1} + I_i + P_i + S_i] +\mathbb{E}[W_i \mathbbm{1}_{I_i + P_i<S_i}] \\ &\hspace{-0.5in}=  \mathbb{E}[P_{i-1} + I_i + P_i + S_i] +\mathbb{E}[W_ie^{-\mu (I_i + P_i)}]  \\ &\hspace{-0.5in}=2\mathbb{E}[P] + \frac{1}{\lambda} + \frac{1}{\mu} + \frac{\lambda}{\mu(\lambda + \mu)}MGF_{\mu}^{(P)}
\end{align}

\subsubsection{$\mathbb{E}[(X_{i}+T_{i})\mathbbm{1}_{i=i^*} \ | \ K_{i-1}=(B)]$}

Conditioned on (B) state observed by packet $i-1$, the next packet index $i$ will be the minimum index among all those being served together only if the next packet $i$ arrives after the residual time $W_{i-1}$. Therefore, if $I_i + P_i < W_{i-1}$, then $\mathbbm{1}_{i=i^*} = 0$ and we have $\mbox{Pr}(i=i^* \ | \ K_{i-1}=(B)) = 1 - \frac{\lambda}{\lambda + \mu} MGF_{\mu}^{(P)}$. In this case, if $I_i + P_i > W_{i-1} + S_{i-1}$, then $T_i = P_i + S_i$. If $W_{i-1} < I_i + P_i < W_{i-1} + S_{i-1}$, then $T_i = P_i + W_i + S_i$ and we have
\begin{align}\nonumber
\mathbb{E}[(X_{i}+T_{i})&\mathbbm{1}_{i=i^*} \ | \ K_{i-1}=(B)] \\ \nonumber &\hspace{-0.5in}= \mathbb{E}[(P_{i-1} + I_i + P_i + S_i)(1-e^{-\mu (I_i + P_i)})] \\ &\ \ +\mathbb{E}[W_i \mu (I_i + P_i)e^{-\mu (I_i + P_i)}]  \\ &\hspace{-0.5in}=2\mathbb{E}[P] + \frac{1}{\lambda} + \frac{1}{\mu} - \frac{(\mathbb{E}[P] + \frac{1}{\mu})\lambda}{\lambda + \mu}MGF_{\mu}^{(P)} 
\end{align}
We use ergodicity of the system to conclude as follows:
\begin{align*}\nonumber
\mathbb{E}[(X_{i}+T_{i})\mathbbm{1}_{i=i^*}] &= 2\mathbb{E}[P] + \frac{1}{\lambda} + \frac{1}{\mu}  \\ &\hspace{-0.8in} + (1- 2p_B)\frac{\lambda MGF_{\mu}^{(P)}}{\mu(\lambda + \mu)} - p_B \frac{\mathbb{E}[P] \lambda}{\lambda + \mu}MGF_{\mu}^{(P)} \end{align*} \begin{align*}
\mbox{Pr}(i=i^*) &= 1 - p_B \frac{\lambda}{\lambda + \mu} MGF_{\mu}^{(P)}
\end{align*}
and finally we have $\mathbb{E}[X_{i^*} + T_{i^*}] = \frac{\mathbb{E}[(X_{i}+T_{i})\mathbbm{1}_{i=i^*}] }{\mbox{Pr}(i=i^*)}$.

\section{Numerical Results and Discussion}
\label{sec:Numres}

In this section, we provide numerical results for AoI with respect to system parameters under various service distributions. Additionally, we performed packet-based queue simulations for $10^6$ packets as verification of all numerical results. Our goal is to obtain the best operating point determined by mean service times $\mathbb{E}[P]$ and $\mathbb{E}[S]=\frac{1}{\mu}$ given that $\frac{1}{\mu}=g(\mathbb{E}[P])$. The fact that $g(.)$ is a monotone decreasing function enables us to trade mean computation time with mean transmission time and there is an optimal operating point with respect to average AoI and average peak AoI. We observe in general that a decrease in average AoI comes at the cost of increased average peak AoI. To understand the tradeoff between average AoI and average peak AoI, we optimize the mean service times with the objective of weighted sum of AoI and average peak AoI for different weights indicating the importance of each.
\begin{align}\label{ths}
\min_{\mathbb{E}[P] \geq 0} \omega_1 \mathbb{E}[\Delta] + \omega_2 \mathbb{E}[PAoI]
\end{align}
For simplicity, we use $g(\mathbb{E}[P])=B_0 e^{-\alpha \mathbb{E}[P]}$. This function is indicative of an exponential decrease in transmission time and is a good fit for applications that require processing data at the transmitter for higher accuracy and transmitting data that remain to be processed. Note that this selection of $g(.)$ is convex and hence the potential improvement obtained by trading computation time with transmission time is expected to be bounded due to diminishing returns. Its smoothness makes it suitable for use as an approximation for many potential non-smooth variations of it. We let the expected processing time to be selected from the interval $P_{min} \leq \mathbb{E}[P] \leq P_{max}$. We take $P_{min}=1$ and $P_{max}=10$ in the rest.

We use Gamma distributed computation time with mean $\mathbb{E}[P]$. In particular, we use the probability density function $f_P(p)=\frac{k^k \kappa^k}{\Gamma(k)}p^{k-1}e^{-k \kappa p}$ for $p \geq 0$ where $\kappa=\frac{1}{\mathbb{E}[P]}$ and $k >0$ determines the variance. In particular, the variance gets larger as $k$ gets smaller. Indeed, this distribution converges to an impulse at $\mathbb{E}[P]$ as $k$ grows large. We have the following closed form expressions for this Gamma distribution:
\begin{align*}
MGF^{(P)}_{\mu}&=\left(1+\frac{\mu }{k \kappa}\right)^{-k} \\  MGF^{(P,1)}_{\mu}&=\frac{1}{\kappa}\left(1+\frac{\mu }{k \kappa}\right)^{-k-1} 
\end{align*}
 
\begin{figure}[!t]
\centering{
\hspace{-0.3cm} 
\includegraphics[totalheight=0.28\textheight]{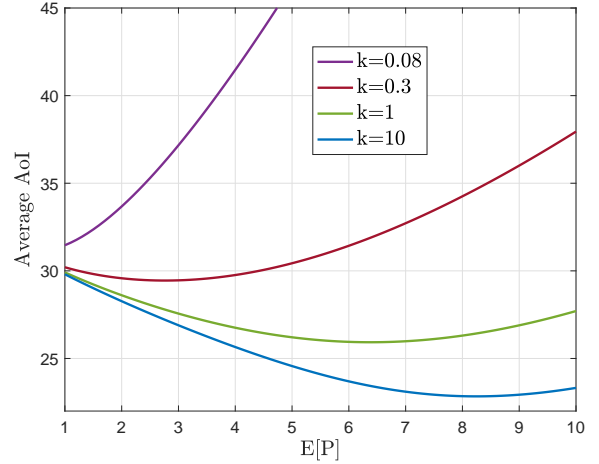}}\vspace{-0.015in}
\caption{\sl Average AoI with respect to $\mathbb{E}[P]$ for fixed $\lambda=0.4$, $B_0=15$ and $\alpha=0.1$.}\vspace{-0.015in}
\label{fig:num3} 
\end{figure}

\begin{figure}[!t]
\centering{
\hspace{-0.3cm} 
\includegraphics[totalheight=0.28\textheight]{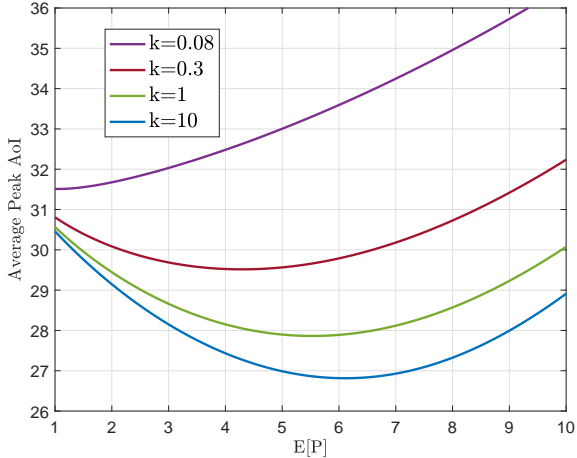}}\vspace{-0.015in}
\caption{\sl Average peak AoI with respect to $\mathbb{E}[P]$ for fixed $\lambda=0.4$, $B_0=15$ and $\alpha=0.1$.}\vspace{-0.02in}
\label{fig:num1} 
\end{figure}

We start with Figs. \ref{fig:num3} and \ref{fig:num1}, where we compare the average AoI and average peak AoI with respect to $\mathbb{E}[P]$ in $[1,10]$ interval under different computing time variances indicated by $k$ (larger $k$ means smaller variance). We observe that as the variance of computing time is decreased, both average AoI and average peak AoI decrease uniformly. This observation supports the usefulness of determinacy in this tandem queue system. Usefulness of determinacy has been observed in the seminal paper \cite{talak2018can} for single server first come first served systems and our work extends this conclusion at least numerically in the tandem queue model with packet management. 

\begin{figure}[!t]
\centering{
\hspace{-0.3cm} 
\includegraphics[totalheight=0.28\textheight]{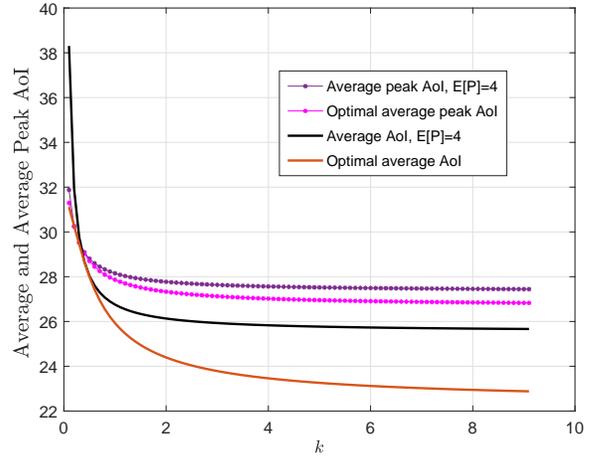}}\vspace{-0.015in}
\caption{\sl Average and average peak AoI with respect to $k$ for fixed $\lambda=0.4$, $B_0=15$ and $\alpha=0.1$.}\vspace{-0.02in}
\label{fig:num2} 
\end{figure}

Next, in Fig. \ref{fig:num2}, we compare the gains in average AoI that could be obtained by judiciously selecting the operating point $\mathbb{E}[P]$. In particular, we select $\mathbb{E}[P]=4$ arbitrarily and compare its performance with optimal selection. We verify once again the monotonic decrease of average and average peak AoI with respect to decreasing variance. Note that the improvement in average AoI is significantly higher for larger variances while it is not the case for peak AoI. This is analogous to the effect of waiting as in \cite{infocom_w, infocom_arxiv} where larger variance in the service time distribution yields a higher improvement in average AoI. In particular, the mean service time $\mathbb{E}[P]$ has an analogous role as \textit{waiting time} from the point of view of the second queue. Additionally, we note that the improvement in average AoI shows a larger margin compared to average peak AoI.

\begin{figure}[!t]
\centering{
\hspace{-0.3cm} 
\includegraphics[totalheight=0.28\textheight]{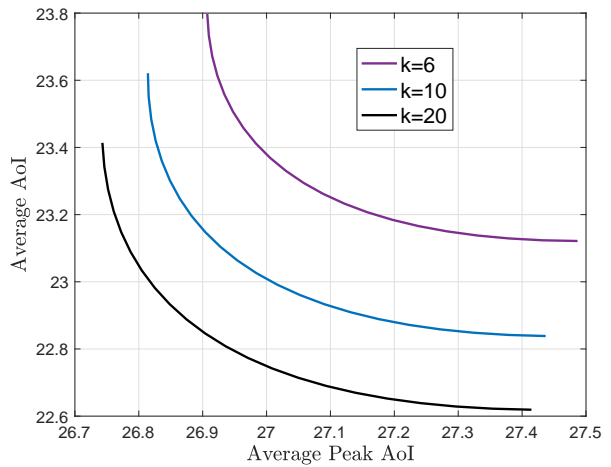}}\vspace{-0.015in}
\caption{\sl Optimal tradeoff curves for average AoI vs. average peak AoI with differing variances and fixed $\lambda=0.4$, $B_0=15$ and $\alpha=0.1$.}\vspace{-0.015in}
\label{fig:num4} 
\end{figure}

In earlier figures, we also observe that average peak AoI could take a smaller value than average AoI for larger variances while for $k \geq 1$ it is always larger. The optimal values of $\mathbb{E}[P]$ are quite different for average AoI and average peak AoI for different $k$. Indeed, in Figs. \ref{fig:num3} and \ref{fig:num1}, we observe that as variance gets larger it is optimal to keep $\mathbb{E}[P]$ at its minimum level $1$ for both optimal average AoI and average peak AoI; however, for smaller variances optimal values for average AoI and peak AoI are different. These indicate that there is a tradeoff between average AoI and average peak AoI. In Fig. \ref{fig:num4}, we plot the optimal tradeoff obtained by solving the weighted optimization in (\ref{ths}) for differing service time variances. In particular, for each $k$ determining the service time variance, we solve (\ref{ths}) for all possible $\omega_1$ and $\omega_2$ and plot all possible operating points as tuples of average AoI and average peak AoI. This characterizes the optimal tradeoff between average AoI and average peak AoI. We observe that this tradeoff becomes more apparent for smaller service time variances.

\begin{figure}[!t]
\centering{
\hspace{-0.3cm} 
\includegraphics[totalheight=0.28\textheight]{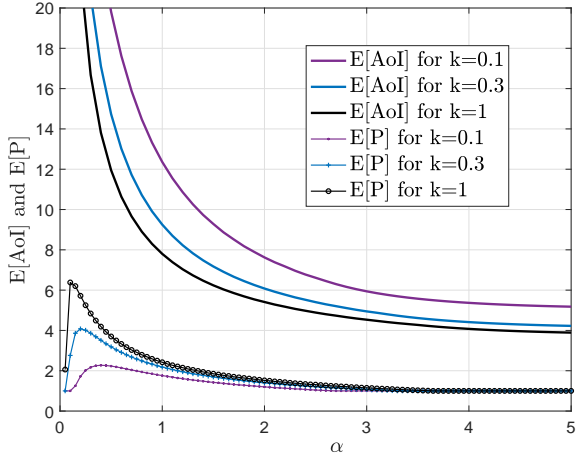}}\vspace{-0.01in}
\caption{\sl Optimal average AoI and optimal $\mathbb{E}[P]$ with respect to $\alpha$ for various $k$ and fixed $\lambda=0.4$ and $B_0=15$.}\vspace{-0.02in}
\label{fig:num5} 
\end{figure}

We finally provide in Fig. \ref{fig:num5} the optimal average AoI and the best selection of $\mathbb{E}[P]$ in $[P_{min},P_{max}]$ that minimizes average AoI plotted with respect to the exponent appearing in $g(.)$ function relating $\mathbb{E}[P]$ and $\mathbb{E}[S]$. The larger $\alpha$ is, the more effective the amount of computation performed in the first queue is and the larger the reduction in service time of the transmission queue is. We observe monotonicity with respect to $\alpha$ and different points of convergence for different $k$ values. We also note that optimizer $E[P]$ has a unimodal shape with respect to $\alpha$. In particular, the extreme values of $\alpha$ requires to set $\mathbb{E}[P]$ to $P_{min}$ whereas larger values of $\mathbb{E}[P]$ achieve better tradeoff between computation and communication in moderate values of $\alpha$. It is also remarkable that optimal $\mathbb{E}[P]$ shows monotonicity with respect to the variance of computation time.    

\section{Conclusions}
\label{sec:Conc}

We considered stationary distribution analysis for average AoI and average peak AoI in a tandem non-preemptive queue with a computation server and a transmission queue with arrivals coming from the computation queue. Moreover, there is a functional dependence between the mean service times of the first and the second queues. We obtain closed form expressions for average peak AoI and average AoI in this system. Our expressions provide explicit relations among the parameters in the system. Our numerical results show the benefits of judiciously determining the operating point and the tradeoff between average peak AoI and average AoI for different computation time distributions.

\end{document}